\documentclass[twocolumn,english,reprint, longbibliography, superscriptaddress, breaklinks=true, showkeys, showpacs=false, nofootinbib]{revtex4}
\usepackage[T1]{fontenc}
\usepackage[latin9]{inputenc}
\usepackage{color}
\usepackage{babel}
\usepackage{amsmath}
\usepackage{amsthm}
\usepackage{amsfonts}
\usepackage{amssymb}
\usepackage{graphicx}
\usepackage{subfigure}
\usepackage{physics}

\usepackage[unicode=true,pdfusetitle,
 bookmarks=true,bookmarksnumbered=false,bookmarksopen=false,
 breaklinks=true,pdfborder={0 0 0},backref=false,colorlinks=true]
 {hyperref} 
\usepackage{breakurl}

\makeatletter
\@ifundefined{textcolor}{}
{%
 \definecolor{BLACK}{gray}{0}
 \definecolor{WHITE}{gray}{1}
 \definecolor{RED}{rgb}{1,0,0}
 \definecolor{GREEN}{rgb}{0,1,0}
 \definecolor{BLUE}{rgb}{0,0,1}
 \definecolor{CYAN}{cmyk}{1,0,0,0}
 \definecolor{MAGENTA}{cmyk}{0,1,0,0}
 \definecolor{YELLOW}{cmyk}{0,0,1,0}
}

\pdfoutput=1
\hypersetup{colorlinks=true,citecolor=blue,linkcolor=cyan,urlcolor=blue,filecolor= green, breaklinks=true}
\usepackage{url}
\usepackage{breakurl}
\makeatother

\begin{document}

\title{The effect of stationary axisymmetric spacetimes in interferometric visibility}

\author{Marcos L. W. Basso}
\email{marcoslwbasso@mail.ufsm.br}
\address{Departamento de F\'isica, Centro de Ci\^encias Naturais e Exatas, Universidade Federal de Santa Maria, Avenida Roraima 1000, Santa Maria, Rio Grande do Sul, 97105-900, Brazil}

\author{Jonas Maziero}
\email{jonas.maziero@ufsm.br}
\address{Departamento de F\'isica, Centro de Ci\^encias Naturais e Exatas, Universidade Federal de Santa Maria, Avenida Roraima 1000, Santa Maria, Rio Grande do Sul, 97105-900, Brazil}

\selectlanguage{english}%

\begin{abstract} 
 In this article, we consider a scenario in which a spin-$1/2$ quanton goes through a superposition of co-rotating and counter-rotating circular paths, which play the role of the paths of a Mach-Zehnder interferometer in a stationary and axisymmetric spacetime. Since the spin of the particle plays the role of a quantum clock, as the quanton moves in a superposed path it gets entangled with the momentum (or the path), and this will cause the interferometric visibility (or the internal quantum coherence) to drop, since, in stationary axisymmetric spacetimes there is a difference in proper time elapsed along the two trajectories.  However, as we show here, the proper time of each path will couple to the corresponding local Wigner rotation, and the effect in the spin of the superposed particle will be a combination of both. Besides, we discuss a general framework to study the local Wigner rotations of spin-$1/2$ particles in general stationary axisymmetric spacetimes for circular orbits.
\end{abstract}


\maketitle

\section{Introduction}

The effects of rotation in space-time are the sources of important and stimulating research as well for the scientific debates which have permeated the field of physics in the last centuries. In the beginning of the 19th century, Sagnac predicted and experimentally verified that there exists a shift of the interference pattern when an interferometric apparatus is rotating, compared to what is observed when the device is at rest \cite{Sagnac, Rug}. General relativity predicts that two freely counter revolving test particles in the exterior field of a central rotating body take different periods of time to complete the same orbit. This time difference leads to the gravitomagnetic clock effect (GCE) \cite{Tarta}. In fact, the influence of the angular momentum of the field also appears in the form of frame dragging of local inertial frames, i.e., the Lense-Thirring effect \cite{Lense}, which was experimentally verified in Refs. \cite{lucchesi,Ciu}. A time difference like in the gravitomagnetic clock effect can be easily converted into a phase difference due to the difference of proper time elapsed in each path. Besides, it is the same mechanism which produces, in special relativity, the Sagnac effect. However, when a gravitational field is present the situation is essentially the same, except for corrections produced by curvature and angular momentum of the rotating body \cite{Tartagl}. Until now, several approaches has been taken to derive and to propose experiments to verify the gravitomagnetic clock effect \cite{Cohen, Gronwald, Tartag, Tart, Bini, Faruque, Faru, Ruggi}. 

In particular, Faruque \cite{Faru} found a quantum analogue of the classical gravitomagnetic clock effect, by presenting an approximation to the solution of the Dirac equation in the Schwarzschild field through the use of Foldy-Wouthuysen Hamiltonian. The analogy consists in the realization that the difference of the periods of oscillation of two states with two different total angular momentum quantum numbers has an analog form of the classical clock effect found in general relativity. However, it is worth pointing out that the Schwarzschild metric represents a static and spherically symmetric spacetime, while the GCE requires a stationary and axisymmetric spacetime. Therefore, in this work, we take another approach to study the effect of stationary and axisymmetric spacetimes in spin-$1/2$ particles. 

Zych \textit{et al.} \cite{Zych} considered a Mach-Zehnder interferometer placed in a gravitational potential, with a `clock' used as an interfering particle, i.e., an evolving internal degree of freedom of a particle. Due to the difference in proper time elapsed along the two trajectories, the `clock' evolves to different quantum states for each path of the interferometer. In this article, we present an analogue scenario in which a spin-$1/2$ quanton goes through a superposition of co-rotating and counter-rotating geodetic circular paths, which play the role of the paths of a Mach-Zehnder interferometer in a stationary and axisymmetric spacetime, since there exists a difference in proper time elapsed along the two trajectories. Here the spin of the particle plays the role of a quantum clock, and as the quanton \cite{Leblond} moves in a superposed path, it gets entangled with the momentum (or the path), and this will cause the interferometric visibility (or the internal quantum coherence) to drop, since, due to the wave-particle duality, the interferometric visibility will decrease by an amount given by the which-way information accessible from the final state of the clock, which gets entangled with the external degree of freedom of the particle. However, as we will show, the proper time of each path will couple to the corresponding local Wigner rotation, and the effect in the spin of the superposed particle will be a combination of both. Thus, our work follows the line of research towards probing general relativistic effects in quantum phenomena \cite{Zych, Magdalena, Brodutch, Costa}. It is worth mentioning that the motion of spinning particles, either classical or quantum, does not follow geodesics because the spin and curvature couple in a non-trivial manner \cite{Papapetrou}. However, the corrections, which are of order $\mathcal{O}(\hbar)$, for quantum particles, become important only when the motion of the particle is ultra-relativistic or/and the source is a supermassive object \cite{Lanzagorta}. In contrast, in this manuscript, we work only in slow-rotating and weak field approximation. The effect of spin-curvature coupling on the interferometric visibility of a spin-1/2 quanton will be considered in a future work. 

To accomplish this task, we use the method developed in Ref. \cite{Terashima} by Terashima and Ueda,  who studied EPR correlations and the violation of Bell's inequalities in the Schwarzschild spacetime, by considering a succession of infinitesimal local Lorentz transformations and how the spin-$1/2$ representations of these local Lorentz transformations (i.e. local Wigner rotations) affect the state of the spin. Besides, we discuss a general framework to study the local Wigner rotations of spin-$1/2$ particles in general stationary axisymmetric spacetimes for circular orbits. The benefit of the approach taken here, through the representation of the local Lorentz transformation, is that one does not need to know the (internal) Hamiltonian of the system and how it couples to the gravitational field, as was the case in Refs. \cite{Faruque, Zych, Costa}. In contrast, given an arbitrary spacetime metric, we only need to calculate the local Wigner rotations, which is a straightforward procedure.

%Its applications are several, such as fiber optic gyroscopes, used in inertial navigation, and ring laser gyroscopes, used in geophysics \cite{Rugg}, as well in  the global positioning system \cite{Ashby}. Besides, exist many experiments that suggest the sagnac effect is universal, in the sense that it's independent of the nature of the interfering beams \cite{Ml}. For instance, the Sagnac effect with matter waves has been verified experimentally using Cooper pair \cite{Zim}, neutrons \cite{Atw}, electrons \cite{Hassel}. In a series of remarkable experiments, Werner et al. \cite{Werner} demonstrated the effect of the terrestrial rotation on neutrons phase.

The organization of this article is as follows. In Sec. \ref{sec:spin}, we review the spin-$1/2$ dynamics in curved spacetimes. In Sec. \ref{sec:inter}, we describe the local Wigner rotations of spin-$1/2$ particles  for circular geodesics in general stationary axisymmetric spacetimes. In Sec. \ref{sec:intvis}, we consider a spin-$1/2$ quanton in a superposition of co-rotating and counter-rotating geodetic circular paths and show how the interferometric visibility is affected by spacetime effects. Thereafter, in Sec. \ref{sec:con}, we give our conclusions.

%-------------------------------
\section{Qubit Dynamics in Curved Spacetimes}
\label{sec:spin}
\subsection{Spin States in Local Frames }
Coordinate transformations in general relativity are described through the group $GLR(4)$, which is the set of all real regular $4 \times 4$ matrices. However, it is well  known that $GLR(4)$ does not have a spinor representation. Therefore, to properly treat spins in the context of curved spacetimes, the use of local frames of reference is required. These frames are defined through an orthonormal basis or tetrad field (or vielbein), which is a set of four linearly independent orthonormal 4-vector fields defined at each point of spacetime  \cite{Wald}. The differential structure of the spacetime, which is a differential manifold $\mathcal{M}$ \cite{Carroll}, provides, in each point $p$, a coordinate basis for the tangent space $T_{p}(\mathcal{M})$, as well as for the cotangent space $T^*_{p}(\mathcal{M})$, given by $\{\partial_{\mu}\}$ and $\{ dx^{\nu} \}$, respectively, such that $dx^{\nu}(\partial_{\mu}) := \partial_{\mu}x^{\nu} = \delta^{\ \nu}_{\mu}$. Therefore, the metric can be expressed as $g = g_{\mu \nu}(x) dx^{\mu} \otimes dx^{\nu}$, and the elements of the metric, which encodes the gravitational field, are given by $g_{\mu \nu}(x) = g(\partial_\mu, \partial_{\nu})$. Since the coordinate basis $\{\partial_{\mu}\} \subset T_p(\mathcal{M})$ and $\{ dx^{\nu} \} \subset T^*_p(\mathcal{M})$ are not necessarily orthonormal, it is always possible set up any basis as we like. In particular, we can form an orthonormal basis with respect to the pseudo-Riemannian manifold (spacetime) on which we are working. Following Ref. \cite{Nakahara}, let us consider the linear combination
\begin{align}
    & e_a = e_a^{\ \mu}(x) \partial_\mu, \ \ \ e^a = e^a_{\ \mu}(x)dx^{\mu}, \\
    & \partial_{\mu} = e^a_{\ \mu}(x)e_a, \ \ \ dx^{\mu} = e_a^{\ \mu}(x)e^a.
\end{align}  
At each point $p \in \mathcal{M}$, The Minkowski metric $\eta_{ab} = \text{diag}(-1,1,1,1)$ in the local frame and the spacetime metric tensor $g_{\mu \nu}(x)$ are related by the tetrad field:
\begin{align}
    & g_{\mu \nu}(x)e_a^{\ \mu}(x)e_b^{\ \nu}(x) = \eta_{ab},\\ &  \eta_{ab}e^a_{\ \mu}(x)e^b_{\ \nu}(x) = g_{\mu \nu}(x) \label{eq:metr},
\end{align}
with
\begin{align}
    e^a_{\ \mu}(x)e_b^{\ \mu}(x) = \delta^{a}_{\ b}, \ \ \ e^a_{\ \mu}(x)e_a^{\ \nu}(x) = \delta_{\mu}^{\ \nu}.
\end{align}
Above and in what follows, Latin letters $a, b, c, d,\cdots$ refer to coordinates in the local frame; Greek indices $\mu, \nu, \cdots$ run over the four general-coordinate labels; and repeated indices are to be summed over. The components of the vielbein transform objects from the general coordinate system $x^{\mu}$  to the local frame $x^a$, and vice versa. Therefore, it can be used to shift the dependence of spacetime curvature of the tensor fields to the tetrad fields. In addition, Eq. (\ref{eq:metr}) informs us that the tetrad field encodes all the information about the spacetime curvature hidden in the metric. Besides, the tetrad field $\{e_a^{\ \mu}(x), a = 0,1,2,3\}$ is a set of four 4-vector fields, which transforms under local Lorentz transformations in the local system. The choice of the local frame is not unique, since the  local frame remains local
under the local Lorentz transformations. Therefore, a tetrad representation of a particular metric is not uniquely defined, and different tetrad fields will provide the same metric tensor, as long as they are related by local Lorentz transformations \cite{Misner}.

By constructing the local Lorentz transformation, we can define a particle with spin-$1/2$ in curved spacetimes as a particle whose one-particle states furnish the spin-$1/2$ representation of the local Lorentz transformation, and not as a state of the diffeomorphism group \cite{Terashima}. Therefore, if $p^{\mu}(x) = m u^{\mu}(x)$ represents the four-momentum of such particle, with $p^{\mu}(x) p_{\mu}(x) = -m^2$, in the general reference frame, then the momentum related to the local frame is given by $p^a(x) = e^a_{\ \mu}(x) p^{\mu}(x)$, where $m$ is the mass of the quanton, $u^{\mu}(x)$ is the four-velocity in the general coordinate system, and we already putted $c = 1$. Thus, in the local frame at point $p \in \mathcal{M}$ with coordinates $x^a = e^a_{\ \mu}(x) x^{\mu}$, a momentum eigenstate of a Dirac particle in a curved spacetime is given by \cite{Lanzagorta}
\begin{align}
 \ket{p^a(x), \sigma; x} := \ket{p^a(x), \sigma; x^{a}, e^a_{\ \mu}(x), g_{\mu \nu}(x)},
\end{align}
and represents the state with spin $\sigma$ and momentum $p^a(x)$ as observed from the position $x^a = e^a_{\ \mu}(x) x^{\mu}$ of the local frame defined by $ e^a_{\ \mu}(x)$ in the spacetime $\mathcal{M}$ with metric $g_{\mu \nu}(x)$. It's noteworthy that the description of a Dirac particle state can only be provided regarding the tetrad field and the local structure that it describes.  By definition, the state $\ket{p^a(x), \sigma; x}$ transforms as the spin-$1/2$ representation under the local Lorentz transformation. In special relativity, a one-particle spin-$1/2$ state $\ket{p^a, \sigma}$ transforms under a Lorentz transformation $\Lambda^{a}_{b}$ as \cite{Weinberg}
\begin{equation}
    U(\Lambda)\ket{p^a, \sigma} = \sum_{\lambda} D_{\lambda \sigma} (W(\Lambda,p)) \ket{\Lambda p^a, \lambda},
\end{equation}
where $D_{\lambda, \sigma}(W(\Lambda,p))$ is a unitary representation of the Wigner's little group, whose elements are Wigner rotations $W^{a}_{b} (\Lambda,p)$ \cite{Eugene}. The subscripts can be suppressed and one can write $U(\Lambda) \ket{p^a, \sigma} = \ket{\Lambda p^a} \otimes D (W(\Lambda, p)) \ket{\sigma},$ as sometimes we will do. In other words, under a Lorentz transformation $\Lambda$, the momenta $p^a$ goes to $\Lambda p^a$, and the spin transforms under the representation $D_{\lambda \sigma}(\Lambda, p)$ of the Wigner's little group \cite{Onuki}. The spin of the particle is then rotated by an amount that depends on the momentum of the particle. In contrast, in general relativity, single-particle states now form a local representation of the inhomogeneous Lorentz group at each point $p \in \mathcal{M}$, thus we have
\begin{equation}
    U(\Lambda(x))\ket{p^a(x), \sigma;x} = \sum_{\lambda} D_{\lambda \sigma}(W(x)) \ket{\Lambda p^a(x), \lambda;x} \label{eq:unit},
\end{equation}
where $W(x) := W(\Lambda(x), p(x))$ is a local Wigner rotation.

%-------------------------
\subsection{Spin Dynamics}
Following Terashima and Ueda \cite{Terashima}, different local  frames can be defined at each point of the curved spacetime. In the local frame at point $p$ with coordinates $x^a = e^a_{\ \mu}(x) x^{\mu}$, the momentum of the particle is given by $p^a(x) = e^a_{\ \mu}(x) p^{\mu}(x)$. After an infinitesimal proper time $d \tau$,  during which the quanton moves to the new point $x'^{\mu} = x^{\mu} + u^{\mu} d\tau$ along its world line, the momentum of the particle in the local frame at the new point becomes $p^a(x') = p^a(x) + \delta p^a(x)$. Such infinitesimal change can be decomposed as
\begin{equation}
    \delta p^a(x) = e^a_{\ \mu}(x) \delta p^{\mu}(x) + \delta e^a_{\ \mu}(x)p^{\mu}(x). \label{eq:momen}
\end{equation}
The variation $\delta p^{\mu}(x)$ in the first term on the right hand side of the last equation is due to an external non-gravitational force 
\begin{align}
     \delta p^{\mu}(x)&  = u^{\nu}(x) \nabla_{\nu} p^{\mu}(x) d\tau = m a^{\mu}(x) d\tau  \\
     & =  - \frac{1}{m}(a^{\mu}(x)p_{\nu}(x) - p^{\mu}(x)a_{\nu}(x))p^{\nu}(x) d\tau, \nonumber 
\end{align}
where it was used the normalization condition for $p^{\mu}(x)$ and the fact that $p^{\mu}(x)a_{\mu}(x) = 0$. Besides, $\nabla_{\nu}$ is the covariant derivative and $a^{\mu}(x):=u^{\nu}(x) \nabla_{\nu} u^{\mu}(x)$ is the acceleration due to a non-gravitational force. Meanwhile, the variation of the tetrad field is due to spacetime geometry effects and it is given by
\begin{align}
    \delta e^a_{\ \mu}(x) & = u^{\nu}(x) \nabla_{\nu}e^a_{\ \mu}(x) d\tau \nonumber \\
    & = - u^{\nu}(x) \omega_{\nu \ b}^{\ a}(x) e^b_{\ \mu}(x)d \tau,
\end{align}
where $\omega_{\nu \ b}^{\ a} := e^{a}_{\ \lambda} \nabla_{\nu} e_{b}^{\ \lambda} = - e_{b}^{\ \lambda} \nabla_{\nu} e^{a}_{\ \lambda} $ is the connection 1-form (or spin connection) \cite{Chandra}. Collecting these results and substituting in Eq. (\ref{eq:momen}), we obtain
\begin{equation}
    \delta p^a(x) = \lambda^{a}_{\ b}(x)p^{b}(x) d\tau \label{eq:momvar}, 
\end{equation}
where
\begin{align}
    \lambda^{a}_{\ b}(x) & = - \frac{1}{m}(a^{a}(x)p_{b}(x) - p^{b}(x)a_{a}(x)) + \chi^{a}_{\ b} \nonumber \\
    & = - (a^{a}(x)u_{b}(x) - u^{a}(x)a_{b}(x)) + \chi^{a}_{\ b} \label{eq:infloc}
\end{align}
with $\chi^{a}_{\ b} :=  - u^{\nu}(x) \omega_{\nu \ b}^{\ a}(x)$. The Eqs. (\ref{eq:momvar}) and (\ref{eq:infloc}) constitute an infinitesimal local Lorentz transformation since,  as the particle moves in spacetime during an infinitesimal proper time interval $d \tau$, the momentum in the local frame transforms as $p^{a}(x) = \Lambda^{a}_{\ b}(x) p^b(x)$ where $\Lambda^{a}_{\ b}(x) = \delta^{a}_{\ b} + \lambda^{a}_{\ b}(x)d \tau$ \cite{Lanzagorta}. In a similar manner, a spin state transforms via a representation of the local Lorentz transformation. In other words, by using a unitary representation of the local Lorentz transformation, the state $\ket{p^a(x), \sigma; x}$ is now described as $U(\Lambda(x)) \ket{p^a(x), \sigma; x}$ in the local frame at the point $x'^{\mu}$, and Eq. (\ref{eq:unit}) expresses how the state of the spin changes locally as the particle moves from $x^{\mu} \to x'^{\mu}$ along its world line. Therefore, one can see that spacetime tells quantum states how to evolve. For the infinitesimal Lorentz transformation, the infinitesimal Wigner rotation is given by $ W^{a}_{\ b}(x) = \delta^{a}_{\ b} + \vartheta^{a}_{\ b} d \tau,$
where $\vartheta^{0}_{\ 0}(x) = \vartheta^{i}_{\ 0}(x) = \vartheta^{0}_{\ i}(x) = 0$ with
\begin{equation}
    \vartheta^{i}_{\ j}(x) = \lambda^{i}_{\ j}(x) + \frac{\lambda^{i}_{\ 0}(x)p_j(x) - \lambda_{j0}(x)p^i(x)}{p^0(x) + m},
\end{equation}
whereas all other terms vanish. In Ref. \cite{Kilian}, the authors provided an explicit calculation of these elements. Hence, the two-spinor representation of the infinitesimal Wigner rotation is then given by
\begin{align}
    D(W(x)) & = I_{2 \times 2} + \frac{i}{4} \sum_{i,j,k = 1}^{3} \epsilon_{ijk} \vartheta_{ij}(x) \sigma_k d \tau \nonumber \\
    & = I_{2 \times 2} + \frac{i}{2} \boldsymbol{\vartheta} \cdot \boldsymbol{\sigma} d\tau, \label{eq:wigner}
\end{align}
where $I_{2 \times 2}$ is the identity matrix, $\{\sigma_k\}_{k = 1}^3$ are the Pauli matrices, and $\epsilon_{ijk}$ is the Levi-Civita symbol. Moreover, the Wigner rotation for a quanton that moves over a finite proper time interval can be obtained by iterating the expression for the infinitesimal Wigner rotation \cite{Terashima}, and the spin-$1/2$ representation for a finite proper time can be obtained by iterating the Eq. (\ref{eq:wigner}):
\begin{equation}
    D(W(x, \tau)) = \mathcal{T}e^{\frac{i}{2}\int_0^{\tau} \boldsymbol{\vartheta} \cdot \boldsymbol{\sigma} d\tau'}, \label{eq:time}
\end{equation}
where $\mathcal{T}$ is the time-ordering operator \cite{Terashima}, since, in general, the Wigner rotation varies at different points along the trajectory.

%-------------------------------
\section{Wigner Rotation for Circular Geodesics in Stationary Axisymmetric Spacetimes}
\label{sec:inter}
In this section, we will describe the local Wigner rotations of spin-$1/2$ particles in general stationary axisymmetric spacetimes for circular geodesics. Perhaps, the most important of such type of spacetime is the Kerr metric \cite{Kerr, Teukolsky}, which represents an exact solution of a stationary axisymmetric spacetime produced by a rotating body of mass $M$ and angular momentum $\mathcal{J}$. However, a rotating frame in Minkowski spacetime will also describe the spacetime as a stationary and axisymmetric one, since the rotation axis defines a special direction. Once again, it is worth pointing out that the motion of spinning particles, either classical or quantum, does not follow geodesics because the spin and curvature couples in a non-trivial manner \cite{Papapetrou}. However, the corrections, which are of order $\mathcal{O}(\hbar)$, become important only when the motion of the particle is ultra-relativistic or/and the source is a supermassive object \cite{Lanzagorta}. Therefore, this effect can be safely ignored in the regime explored in this article. Finally, the description of local Wigner rotation for circular non-geodetic orbits is easily extended within the framework developed in the Sec. \ref{sec:spin}, as argued in the end of this section. 

A stationary spacetime is characterized by the fact that there exists a one-parameter group of isometries whose orbits are timelike curves, which implies the existence of a timelike Killing vector field. Analogously, a spacetime is axisymmetric if there exists a one-parameter group of isometries whose orbits are closed spacelike curves, which implies the existence of a spacelike Killing vector field whose integral curves are closed. Thus, a spacetime is stationary and axisymmetric if it possesses both of these symmetries and the actions of both of these groups commute, i.e., the rotations commute with the time translations \cite{Wald}. Therefore, we can choose a coordinate system with $x^0 = t$ and $x^3 = \phi$ such that the metric is independent of them. The general form of a stationary axisymmetric metric is given by \cite{Hobson}
\begin{equation}
    ds^2 = g_{tt}dt^2 + 2g_{t \phi} dt d\phi + g_{rr}dr^2  + g_{\theta \theta}d\theta^2 + g_{\phi \phi}d\phi^2,
\end{equation}
with $g_{\mu \nu} = g_{\mu \nu}(r, \theta)$ and $g_{tt} < 0$. Besides, the metric above can be rewritten in two different ways, with one of them being
\begin{align}
    ds^2  =& (g_{tt} - \omega^2 g_{\phi \phi})dt^2 + g_{rr}dr^2  + g_{\theta \theta}d\theta^2  \nonumber \\
    &+ g_{\phi \phi}(d \phi - \omega dt)^2 \label{eq:metric},
\end{align}
where $\omega = - g_{t \phi}/g_{\phi \phi}$. The form of the metric given by Eq. (\ref{eq:metric}) is useful to define world lines of locally non-rotating observers which carries an orthonormal frame with himself, i.e., it is possible to define the following tetrad fields
\begin{align}
     e^{0}_{\ t} =  & \sqrt{-(g_{tt} -  \omega^2 g_{\phi \phi})}, \ \ e^{1}_{\ r} = \sqrt{g_{rr}}, \ \  e^{2}_{\ \theta} = \sqrt{g_{\theta \theta}}, \nonumber \\
    & e^{3}_{\ t} = -\sqrt{g_{\phi \phi}} \omega, \ \ \ \ e^{3}_{\ \phi} = \sqrt{g_{\phi \phi}},
\end{align}
with all the other components being zero. Also, only the nonzero components will be shown from now on. The inverse of these elements are given by
\begin{align}
    & e_{0}^{\ t} = 1/ \sqrt{-(g_{tt} -  \omega^2 g_{\phi \phi})}, \ \ e_{0}^{\ \phi} = \omega/\sqrt{-(g_{tt} -  \omega^2 g_{\phi \phi})}, \nonumber \\ 
    & e_{1}^{\ r} = 1/\sqrt{g_{rr}}, \ \ e_{2}^{\ \theta} = 1/\sqrt{g_{\theta \theta}}, \ \  e_{3}^{\ \phi} =  1/\sqrt{g_{\phi \phi}},
\end{align}
with $u^{\mu} = e_{0}^{\ \mu}$ being the four-velocity of the local non-rotating observer in the coordinate frame. It is important to realize that this observer is rotating in respect to the coordinate frame. This type of observer is called ZAMO observer \cite{Bardeen}. Meanwhile, the other useful form of the metric is given by
\begin{align}
    ds^2  =& g_{tt}(dt - \alpha d\phi)^2 + g_{rr}dr^2  + g_{\theta \theta}d\theta^2  \nonumber \\
    & + (g_{\phi \phi}- \alpha^2 g_{tt})d\phi^2 \label{eq:metric1},
\end{align}
from which it is possible to define `hovering observers' \cite{Salgado} from the tetrad fields
\begin{align}
     e^{0}_{\ t} =  & \sqrt{-g_{tt}}, \ \ e^{0}_{\ \phi} = -  \sqrt{-g_{tt}}\alpha, \ \ e^{1}_{\ r} = \sqrt{g_{rr}}, \nonumber\\
    & e^{2}_{\ \theta} = \sqrt{g_{\theta \theta}}, \ \  e^{3}_{\ \phi} = \sqrt{g_{\phi \phi} - \alpha^2g_{tt}},
\end{align}
while the inverse non-null components are given by:
\begin{align}
    & e_{0}^{\ t} =  1/\sqrt{-g_{tt}}, \ \  e_{1}^{\ r} = 1/\sqrt{g_{rr}}, \ \ e_{2}^{\ \theta} = \sqrt{g_{\theta \theta}}, \\
    &  e_{3}^{\ t} = \alpha/ \sqrt{g_{\phi \phi} - \alpha^2g_{tt}}, \ \   e_{3}^{\ \phi} = 1/\sqrt{g_{\phi \phi} - \alpha^2g_{tt}}, \nonumber 
\end{align}
where $\alpha = - g_{t \phi}/g_{tt}$. Since we can define $e_{0}^{\ \mu}$ as the four-velocity of the hovering observer, one can see that such observer stands still with respect to the coordinate frame. It is worth mentioning that both local observers are not inertial, as one can see by calculating the four-acceleration \cite{Misner}. 

Since we shall deal only with circular geodesic orbits for massive particles, let us restrict ourselves to the equatorial plane characterized by $\theta = \pi/2$. In this case, the non-vanishing metric components are functions only of $r$, i.e., $g_{\mu \nu} = g_{\mu \nu}(r)$ and $d \theta = 0$. Because of the time independence and axial symmetry of the metric, the energy per unit mass $K$ and the angular momentum per unit mass $J$ about the axis of symmetry are constants of the motion. From the Lagrangian $\mathcal{L} = \frac{1}{2}g_{\mu \nu}\Dot{x}^{\mu}\Dot{x}^{\nu}$ and the normalization condition $u^{\mu}u_{\mu} = -1$, the equations of motion for a massive particle confined to the equatorial plane are
\begin{align}
    & \Dot{t} = - g^{tt}(K - \omega J), \label{eq:dott}\\
    & \Dot{\phi} = g^{-1}_{\phi \phi} J - \omega g^{tt}(K - \omega J), \label{eq:dotphi}\\
    & g_{rr}\Dot{r}^2 = - g^{tt}(K - \omega J)^2 - (g^{\phi \phi} -  \omega^2 g^{tt}) J^2 - 1,
\end{align}
where $\dot{x}^{\mu} = dx^{\mu}/d\tau = u^{\mu}$. By defining $Q(r, K, J) = g_{rr}\Dot{r}^2$, the circular geodesics of radius $r_c$ are characterized by the following conditions 
\begin{align}
    Q(r, K, J)|_{r = r_c} =  0 \ \  \text{and} \ \   \partial_r  Q(r, K, J)|_{r = r_c} = 0.
\end{align}
For a given value of $r$, these are two simultaneous equations for the energy K and angular momentum J that can be solved. In addition, $\partial^2_r  Q(r, K, J)|_{r = r_c} = 0$ gives the innermost circular orbit (ISCO). Therefore, for each circular orbit we have two specific values of $K$ and $J$, which characterizes the co-rotating and counter-rotating orbits \cite{Barden}. Since the values of $K$ and $J$ are determined, for each circular orbit, from the conditions above, $\Dot{t}$ and $\Dot{\phi}$ are also settled for each circular orbit such that the angular velocity of the particle with respect to the coordinate frame is given by
\begin{align}
    \Omega = \frac{\Dot{\phi}}{\Dot{t}} = \omega - \frac{J}{g^{tt} g^{\phi \phi}(K - \omega J)},
\end{align}
which also possesses two specific values for each orbit, corresponding to the co-rotating and counter-rotating motion. One can see this by considering the geodesic equation for $r$ and imposing $\dot{r} = \ddot{r} = 0$ \cite{Ryan}. Thus
\begin{align}
    \Omega_{\pm} = \frac{- \partial_r g_{t\phi} \pm \sqrt{(\partial_r g_{t \phi})^2 - \partial_r g_{t t}\partial_r g_{\phi \phi}}}{\partial_r g_{\phi \phi}}, \label{eq:angvel}
\end{align}
where the $+(-)$ refers to the co-rotating (counter-rotating) circular orbits. Besides, it is important to note that $\Omega_{\pm}$ is only a function of $r$ and the properties of the rotating source. Since a circular orbit also implies that $\dot{\phi} = \Omega \dot{t}$, solving for $\dot{t}$ in the normalization condition $u^{\mu}u_{\mu} = -1$ with $\dot{r} = 0$, and substituting in Eqs. (\ref{eq:dott}) and  (\ref{eq:dotphi}) gives
\begin{align}
    & K_{\pm} = \frac{- g_{tt} - g_{t\phi} \Omega_{\pm}}{\sqrt{- g_{tt} - 2 g_{t\phi} \Omega_{\pm} - g_{\phi \phi} \Omega_{\pm}^2}}, \\
    & J_{\pm} = \frac{g_{t\phi} + g_{\phi\phi} \Omega_{\pm}}{\sqrt{- g_{tt} - 2 g_{t\phi} \Omega_{\pm} - g_{\phi \phi} \Omega_{\pm}^2}}.
\end{align}

The four-velocity in both local frames is given by
\begin{align}
    & u^0 = e^0_{\ \mu} u^{\mu} = e^0_{\ t} u^t + e^0_{\ \phi} u^{\phi}, \\
    & u^3 = e^3_{\ \mu} u^{\mu} = e^3_{\ t} u^t + e^3_{\ \phi} u^{\phi}.
\end{align}
For instance, if we choose the tetrad field corresponding to the hovering observer, then $e^3_{\ t} = 0$. From Eq. (\ref{eq:infloc}), one can see that, for any type of geodesic motion, the change of the tetrad field is equal to the local Lorentz transformations. In particular, for the hovering observer in circular geodesics, the non-zero local Lorentz transformations are expressed by
\begin{align}
    \lambda^0_{\ 1} = -u^t e_1^{\ r}\Big( \Gamma^t_{t r}e^0_{\ t} + \Gamma^{\phi}_{t r}e^0_{\ \phi}\Big) -u^{\phi} e_1^{\ r}\Big( \Gamma^t_{\phi r}e^0_{\ t} + \Gamma^{\phi}_{\phi r}e^0_{\ \phi}\Big), \label{eq:llt} \\
    \lambda^1_{\ 3} = -u^t e^1_{\ r}\Big( \Gamma^r_{t t}e_3^{\ t} + \Gamma^{r}_{t \phi}e_3^{\ \phi}\Big) -u^\phi e^1_{\ r}\Big( \Gamma^r_{\phi t}e_3^{\ t} + \Gamma^{r}_{\phi \phi}e_3^{\ \phi}\Big),  \label{eq:lltt}
\end{align}
where $\Gamma^{\alpha}_{\mu \nu} = \frac{1}{2}g^{\alpha \sigma}(\partial_{\mu} g_{\sigma \nu} + \partial_{\nu} g_{\mu \sigma} - \partial_{\sigma} g_{\mu \nu})$ are Christoffel symbols and $\lambda^0_{\ 1} = \lambda^1_{\ 0}, \lambda^1_{\ 3} = - \lambda^3_{\ 1}$. Thus, one can see that the local Lorentz transformation consists of a boost in the $1$ or $r$-axis directions and rotations about the $2$ or $\theta$-axis. The infinitesimal Wigner rotation associated with the rotation over the 2-axis is given by
\begin{align}
        \vartheta^{1}_{\ 3} = \lambda^{1}_{\ 3} + \frac{\lambda^{1}_{\ 0}u_3}{u^0+ 1}. \label{eq:lwr}
\end{align}

 The difference between $ \vartheta^{1}_{\ 3}$  and $\lambda^{1}_{\ 3}$ gives rise to a precession of the spin, which is the general relativistic version of the Thomas precession \cite{Terashima}. In addition, it is noteworthy that $\vartheta^{1}_{\ 3}$ is only a function of $r$ and of the properties of the rotating source. After the test particle has moved in the circular orbit across some proper time $\tau$, the total angle is given by \begin{align}
    \Theta & = \int \vartheta^{1}_{\ 3} d \tau = \vartheta^{1}_{\ 3} \int \frac{d \tau}{d \phi} d\phi = \vartheta^{1}_{\ 3} \frac{d \tau}{d \phi} \Phi,
\end{align}
 since, for circular orbits, $r$ is fixed and, therefore, $\vartheta^{1}_{\ 3} , d \tau / d \phi$ are constant. The angle $\Phi$ is the angle traversed by the particle during the proper time $\tau$. For instance, in Ref. \cite{Salgado} the local Wigner rotations in the Kerr spacetime geometry were studied. Besides, the same analysis can be done if one chooses the ZAMO observers. 
 
  It is noteworthy that the angle $\Theta$ reflects all the rotations suffered by the spin of the qubit as it moves in the circular orbit, which means that there are two contributions: The ``trivial rotation'' $\Phi$ and the rotation due to gravity \cite{Terashima}. Therefore, to obtain the Wigner rotation angle that is produced solely by spacetime effects, it is necessary to compensate for the trivial rotation angle $\Phi$, i.e., $\varphi := \Theta - \Phi$ is the total Wigner rotation of the spin exclusively due to the spacetime curvature. Besides, the local Wigner rotation, $\vartheta^{1}_{\ 3}$, as well as the proper time along the geodesic will depend if the particle is co-rotating or counter-rotating with respect to the source. By denoting $\tau_+$ the proper time of the particle along the co-rotating geodesic and $\tau_-$ the proper time of the particle along the counter-rotating geodesic, and analogously for the azimuthal angle $\phi$, we have
 \begin{align}
     \frac{d \tau_{\pm}}{d \phi_{\pm}} &= \sqrt{- g_{tt}(\frac{d t_{\pm}}{d \phi_{\pm}})^2 - 2 g_{t\phi}(\frac{d t_{\pm}}{d \phi_{\pm}}) - g_{\phi \phi} } \nonumber \\
     & = \frac{1}{\Omega_{\pm}} \sqrt{-g_{tt} - 2 g_{t\phi}\Omega_{\pm} - g_{\phi \phi}\Omega^2_{\pm}}, \label{eq:tauphi}
 \end{align}
where $\Omega_{\pm} = d \phi_{\pm}/d \tau_{\pm}$. For one closed orbit, we obtain the integrated proper time as
\begin{align}
    \tau_{\pm} = \pm \frac{ 2 \pi}{\Omega_{\pm}} \sqrt{ - g_{tt} - 2 g_{t\phi}\Omega_{\pm} - g_{\phi \phi}\Omega^2_{\pm}},
\end{align}
since $\tau_+ \neq \tau_-$, the difference $\tau_+ - \tau_-$ gives rise to the so called gravitomagnetic clock effect. The name comes from the fact that, in the weak field approximation, the gravitational field may be decomposed into a gravitoelectric field and a gravitomagnetic field in analogy with electromagnetism \cite{Tartaglia}. It is worth emphasize that this effect presupposes that the time difference of the two clocks is taken with respect to a fixed angle $\phi$, i.e., after each clock has covered an azimuthal interval of $2 \pi$. However, by taking into account the Wigner rotation, the proper time of each path $d\tau_{\pm}$ will couple to the corresponding infinitesimal Wigner rotation $\vartheta^{1}_{\ 3}(\pm)$. For instance, if we consider the quanton in a superposed path corresponding to a co-rotating and counter-rotating circular geodesic, then the effect of the total Wigner rotation in the spin of the quanton will not be only due to the difference of proper time elapsed in each path, but actually a difference taking into account the infinitesimal Wigner rotation: $\vartheta^{1}_{\ 3}(+) \tau_+ - \vartheta^{1}_{\ 3}(-)\tau_-$. As noticed in Ref. \cite{Costa}, a `clock' (in our case the quanton's spin) with a finite dimensional Hilbert space has a periodic time evolution which causes periodic losses and rebirths of the visibility with increasing time dilation between the two arms of the interferometer. For a clock implemented in a two-level system (which evolves between two mutually orthogonal states), the visibility will be a cosine function. Therefore, for stationary and axisymmetric spacetimes, the argument of the cosine function will be given by the difference $\vartheta^{1}_{\ 3}(+) \tau_+ - \vartheta^{1}_{\ 3}(-)\tau_-$.

Finally, it is possible to extend this procedure for non-geodetic circular orbits. However, in order for the particle to maintain such non-geodetic circular orbit, it is necessary to apply an external radial force against gravity, allowing the quanton to travel in the circular orbit with a specific angular velocity at a given distance $r$ from the source. The analysis above would still be valid in this case, except for the fact that in this case $\chi^{a}_{\ b} \neq \lambda^{a}_{\ b}$, as one can see from Eq. (\ref{eq:infloc}). For instance, one can use this approach to study the Sagnac effect for spin-$1/2$ particles in circular orbits.

%------------------
\section{Interferometric Visibility in an Earth-like Environment}
\label{sec:intvis}
\begin{figure}[t]
\centering
\includegraphics[scale=0.6]{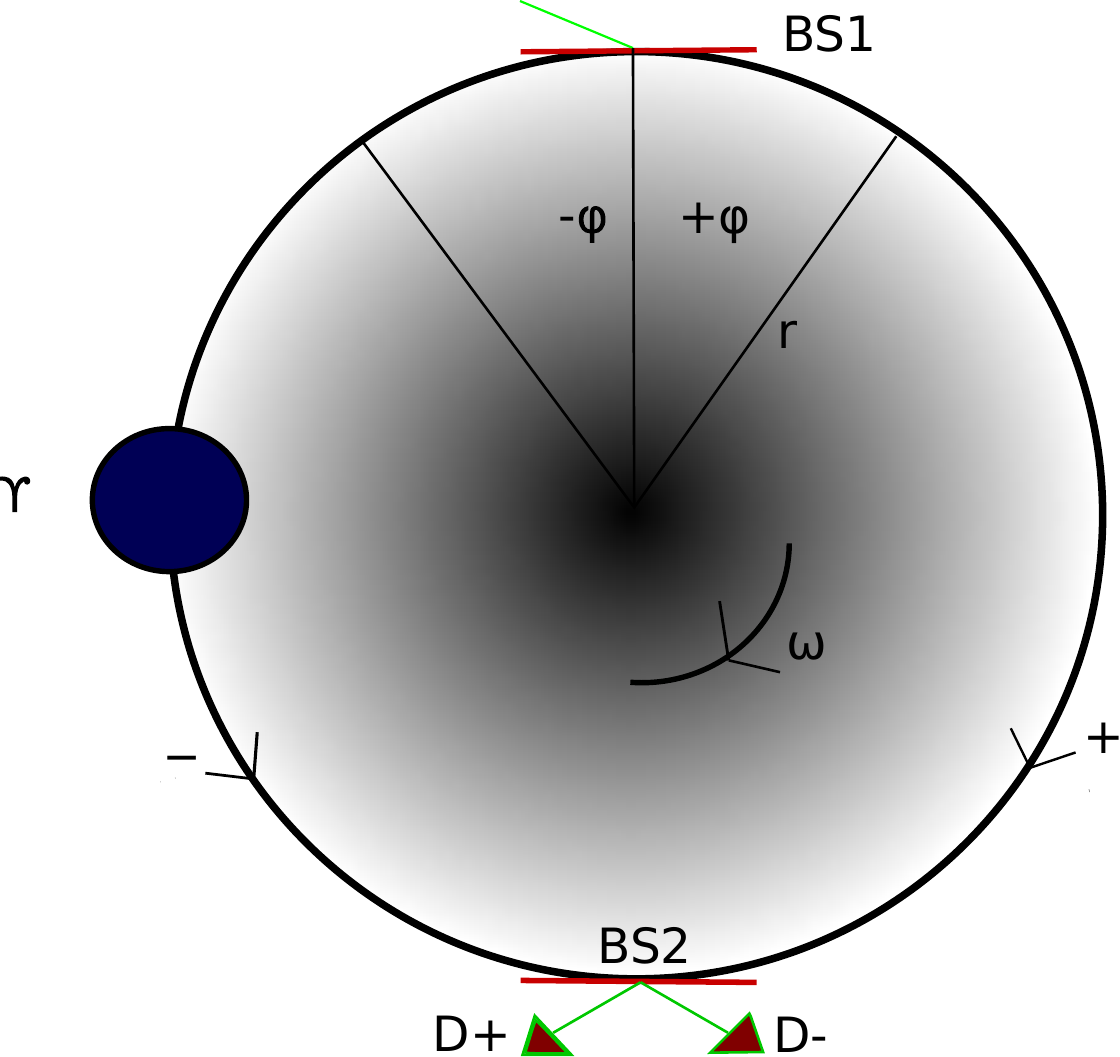}
\caption{A spin-$1/2$ particle in an `astronomical' Mach-Zehnder interferometer. The setup consists of two beam splitters $BS1$ and $BS2$, a phase shifter, which gives a controllable phase $\Upsilon$ to the counter-rotating path $-$, and two detectors $D \pm$. The paths consist of a co-rotating and a counter-rotating circular geodesic centered in a stationary and axisymmetric body of mass $M$ and angular momentum $\mathcal{J}$.}
\label{fig:machrot}
\end{figure}

In this section, we will study the behavior of interferometric visibility under local Wigner rotations as a spin-$1/2$ quanton goes through a superposition of co-rotating and counter-rotating geodetic circular paths, which play the role of the paths of a Mach-Zehnder interferometer in stationary and axisymmetric spacetimes given by an Earth like environment, as depicted in Fig. \ref{fig:machrot}.  To make our investigation easier, we began by assuming that momenta can be treated as discrete variables, as in Refs. \cite{Zych, Terashima}. This can be justified once we can consider narrow distributions centered around different momentum values such that it is possible to represent them by orthogonal state vectors. For instance, if we consider narrow Gaussian distributions for each path, it is a reasonable assumption to assume that both distributions do not overlap since the separation between the paths is bigger than the variance of the Gaussian distributions. Therefore, both distributions are distinguishable and centered around different momentum values such that it is possible to represent them by orthogonal state vectors. 

As well, as pointed out in Ref. \cite{Nasr}, it is possible to assume that the quanton wave-packet has a mean centroid that, in a semiclassical approach, one can regard to describe the motion of the center of mass in each superposed path. The momentum of the particle can be assumed to be distributed properly around the momentum of the centroid. Assuming that the centroid moves along a specified path $x^{\mu}(\tau)$ in the curved spacetime. So, the four-momentum of the centroid as measured in a local frame will be $p^a(x) = e^a_{\ \mu}(x)p^{\mu}(x)$. Therefore, the local observer, defined in each point of the circular geodesic by the tetrad field, deals only with the centroid, i.e., the mean value of the four-momentum which is connected to the four-velocity of the corresponding circular geodesic, and with the centroid of the center of mass in the position basis which is also a Gaussian distribution with the mean value corresponding to the coordinates $x^a$ of the circular geodesic. Therefore, our calculation of the Wigner rotation in this section is with regard to these two mean values, even though both wave-packets have some width. However, when considering the full wave-packet, we have to deal with congruence of geodesics, which is also not a problem, since the wave-packets of momenta and position are not distorted along the geodesic, and therefore the mean value of the position and the momenta remains in the same geodesic (which is part of a geodesic congruence). 

To show this, it is enough to work in the $\hbar^{0}$ order of the WKB expansion for the four-spinor $\psi(x)$ of the Dirac equation in curved spacetime \cite{Lanzagorta}.  In order $\hbar^0$, the Dirac equation in curved spacetimes gives $(\gamma^{\mu}(x) \partial_{\mu} S(x) + m)\psi_0 (x) = 0,$ which is a homogeneous system of four algebraic equations and therefore have non-trivial solutions if and only if $\det[\gamma^{\mu}(x) \partial_{\mu} S(x) + m] = 0$, implying that $m^2 = - \partial^{\mu} S(x) \partial_{\mu}S(x)$. Thus, it is straightforward to see that $p^{\mu} = \partial^{\mu} S(x)$. Therefore, at zero-th order in $\hbar$, it is possible to identify $S(x)$ with the classical action for a particle in a gravitational field: $S(x) = \int p_{\mu} dx^{\mu}$, which also corresponds to the phase of a quantum mechanical particle in curved spacetime \cite{Lanzagorta}. These equations define a family of integral curves $x^{\mu}(\tau)$ of $u^{\mu}$ which is the four-velocity vector field (which in our case will be circular geodesics). Therefore, the computation of $S(x)$ at this approximation order can be accomplished by assuming that the Dirac particle is a classical particle moving with momentum $p^{\mu}$ in curved spacetime. The associated four-velocity field $u^{\mu}$ can be obtained by solving the geodesic equations for a classical test particle in the specific metric under consideration, as we did in the previous section. However, as the extended state of the quantum particle takes on a path superposition, it should be understood that $u^{\mu}$ actually represents a geodesic congruence.  

Besides, it can be shown that the general zeroth order solution to the Dirac equation in curved spacetime $\psi_0(x)$ can be written as an envelope function times a normalized spinor $\psi_0(x) = f (x) \phi_0(x)$, where the envelope function $f(x)$ (which expresses the form of the wave-packet) satisfies the equation $\nabla_{\mu}(f^2(x) u^{\mu}) = 0$, which represents the propagation equation for the envelope function \cite{Palmer}. If we assume that the divergence of the velocity field $u^{\mu}$ vanishes (since we are dealing with geodesics), then 
\begin{align}
    u^{\mu} \nabla_{\mu} f(x) = \frac{d}{d \tau} f(x) = 0.
\end{align}
Thus, the shape of the envelope in the local frame remains unchanged during the evolution. However, of course, because of the uncertainty principle, if the wave packet has finite spatial extent it cannot simultaneously have a sharp momentum, and therefore the divergence in velocity cannot be exactly zero. Nevertheless, we can relax our assumption since, for geodesic congruences, $\nabla_{\mu} u^{\mu}$ measures the rate of change of the local-frame volume \cite{Wald} and assume that the expectation value of $\nabla_{\mu} u^{\mu}$ is small enough:
\begin{align}
    \expval{\nabla_{\mu} u^{\mu}} <<\frac{1}{\tau_{\Gamma}},
\end{align}
where $\tau_{\Gamma}$ is the proper time along some geodesic $\Gamma$. Therefore, we can assume that our initial localized wave-packet is rigidly transported along the geodesic congruence with the mean values of position and momentum following the same geodesic \cite{Palmer}. This discussion is enough to validate the following calculations.

Let us consider the situation in the surroundings of our planet (or the sun). The appropriate (approximate) metric is the one corresponding to a weak axisymmetric field, and it is given by \cite{Tartaglia}
\begin{align}
 g_{tt} & =  - f(r), \ \ g_{rr} = f^{-1}(r), \ \ g_{\theta \theta} \simeq r^2, \\ \nonumber
 & g_{t \phi} \simeq - \frac{r_s a}{r}\sin^2 \theta, \ \ g_{\phi \phi} \simeq r^2 \sin^2 \theta,
\end{align}
where $f(r) = 1 - r_s/r$, with $r_s = 2GM$ being the Schwarzschild radius, and $a = \mathcal{J}/M$, with $\mathcal{J}$ being the angular momentum and $M$ the mass of the rotating body. One can see that, for $\mathcal{J} = 0$, the metric reduces to the Schwarzschild metric. Besides,  we use the Schwarzschild-like coordinates $(t, r, \theta, \phi)$ and assume that the angular momentum is orthogonal to the equatorial plane $\theta = \pi/2$. The Earth-like body is assumed to be spherical and the low rotating approximation is given by the term containing $\mathcal{J}$, which, for instance, in the terrestrial environment is six orders of magnitude smaller than the mass terms \cite{Tartaglia}. 

By restricting ourselves to the equatorial plane, $d \theta = 0$, the relevant components of tetrad field corresponding to hovering observers are given by
\begin{align}
    e^0_{\ t} = \sqrt{f(r)}, & \ \  e^0_{\ \phi} = \frac{r_s a}{r\sqrt{f(r)}}, \ \ e^1_{\ r} = 1/\sqrt{f(r)}, \nonumber \\
    & e^3_{\ \phi} = \sqrt{r^2 + r^2_s a^2/r^2 f(r)},
\end{align}
and all the other components are zero. The inverse of these elements are given by
\begin{align}
        & e_0^{\ t} = 1/\sqrt{f(r)}, \ \  e_3^{\ t} = \frac{- r_s a/rf(r)}{\sqrt{r^2 + r^2_s a^2/r^2 f(r)}} \\ 
        & e_1^{\ r} = \sqrt{f(r)}, \ \  e_3^{\ \phi} = 1/\sqrt{r^2 + r^2_s a^2/r^2 f(r)}.
\end{align}
At each point, the $1-$ and $2-$axis are parallel to the $r$ and $\theta$ directions, respectively. In particular, we choose this tetrad field because the gravitomagnetic effect assumes that the time difference of the two clocks is taken with respect to a fixed angle $\phi$, and therefore hovering observers are necessary in each point.

Now, let us consider the case of a free-falling test spin-$1/2$ particle moving around the source of the gravitational field in a superposition of a co-rotating and counter-rotating geodetic circular orbit, which play the role of the paths of a Mach-Zehnder interferometer, as depicted in Fig. \ref{fig:machrot}. The four-velocity of these circular geodesics in the equatorial plane are given by:
\begin{align}
    & u^t_{\pm} = \frac{1}{\sqrt{- g_{tt} - 2 g_{t\phi} \Omega_{\pm} - g_{\phi \phi} \Omega_{\pm}^2}}, \ \ u^r = 0,  \label{eq:ut}\\
    & u^{\phi}_{\pm} = \frac{\Omega_{\pm}}{\sqrt{- g_{tt} - 2 g_{t\phi} \Omega_{\pm} - g_{\phi \phi} \Omega_{\pm}^2}}, \ \ u^{\theta} = 0, \label{eq:uphi}
\end{align}
 where the angular velocity of the particle with respect to the coordinate frame is obtained directly from Eq. (\ref{eq:angvel}), and is given by
 \begin{align}
     \Omega_{\pm} = - \frac{r_s a}{2 r^3} \pm \sqrt{\Big(\frac{r_s a}{2 r^3}\Big)^2 + \frac{r_s}{2 r^3}},
 \end{align}
where the plus sign refers to the co-rotating orbit, while the minus sign refers to the counter-rotating path, since the angular velocity of the source is given by $\omega = - g_{t \phi}/g_{\phi \phi} = r_s a/r^3$. Ignoring the term $g_{\phi \phi}$ of order $\Omega^2_{\pm}$ in Eq. (\ref{eq:tauphi}), and integrating over $\phi_{\pm}$ \cite{Ryder}, we have
\begin{align}
    \tau_{\pm} & \approx \frac{\Phi_{\pm}}{\Omega_{\pm}} \sqrt{1 - \frac{r_s}{r} + 2 \frac{r_s a}{r}\Omega_{\pm}} \nonumber \\
    & \approx \frac{\Phi_{\pm}}{\Omega_{\pm}}\Big( 1 - \frac{r_s}{2r} + \frac{r_s a}{r}\Omega_{\pm} \Big),
\end{align}
where $\Phi_{\pm} = \pm \Phi$, which implies that the difference of proper time between the two paths is given by
\begin{align}
\tau_+ - \tau_- & = \frac{\Omega_- + \Omega_+}{\Omega_+ \Omega_-}\Big(1 - \frac{r_s}{r}\Big) \Phi + 2 \frac{2 r_s a}{r}\Phi \nonumber \\
& = 2\Phi a.
\end{align}
For one complete revolution, $\tau_+ - \tau_-  = 4 \pi a =  4 \pi J/M$, which is a result independent from both
the radius of the orbit and the gravitational constant $G$ \cite{Ruggiero}. In the terrestrial environment, this proper time difference is of order $10^{-7}s$ \cite{Gronwald, Tart}. Besides, it is possible to derive the same result from the Kerr metric without any approximation (see Ref. \cite{Cohen}).

\begin{figure}[t]
    \centering
    \subfigure[$\lambda^{0}_{\ 1}, \lambda^{1}_{\ 3}$ as a function of $r$, for the parameters $r_s =3$ and $a = 0.1$, with $r \ge 1.4604 r_s$.]{{\includegraphics[scale = 0.57]{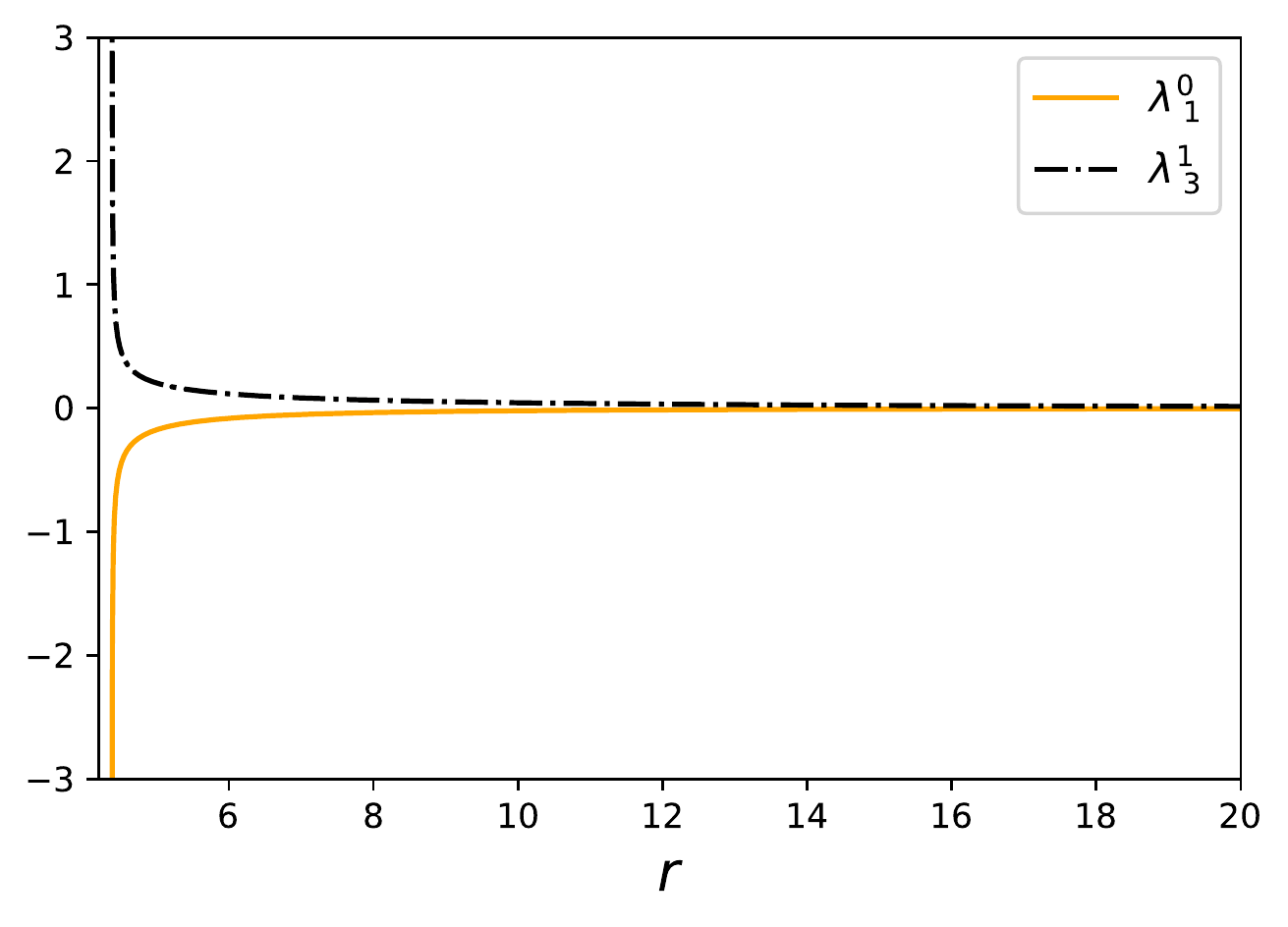}{\label{fig:e}} }}
    \qquad
    \subfigure[$\vartheta^{1}_{\ 3}$ as a function of $r$, for the parameters $r_s =3$ and $a = 0.1$, with $r \ge 1.4604 r_s$.]{{\includegraphics[scale = 0.57]{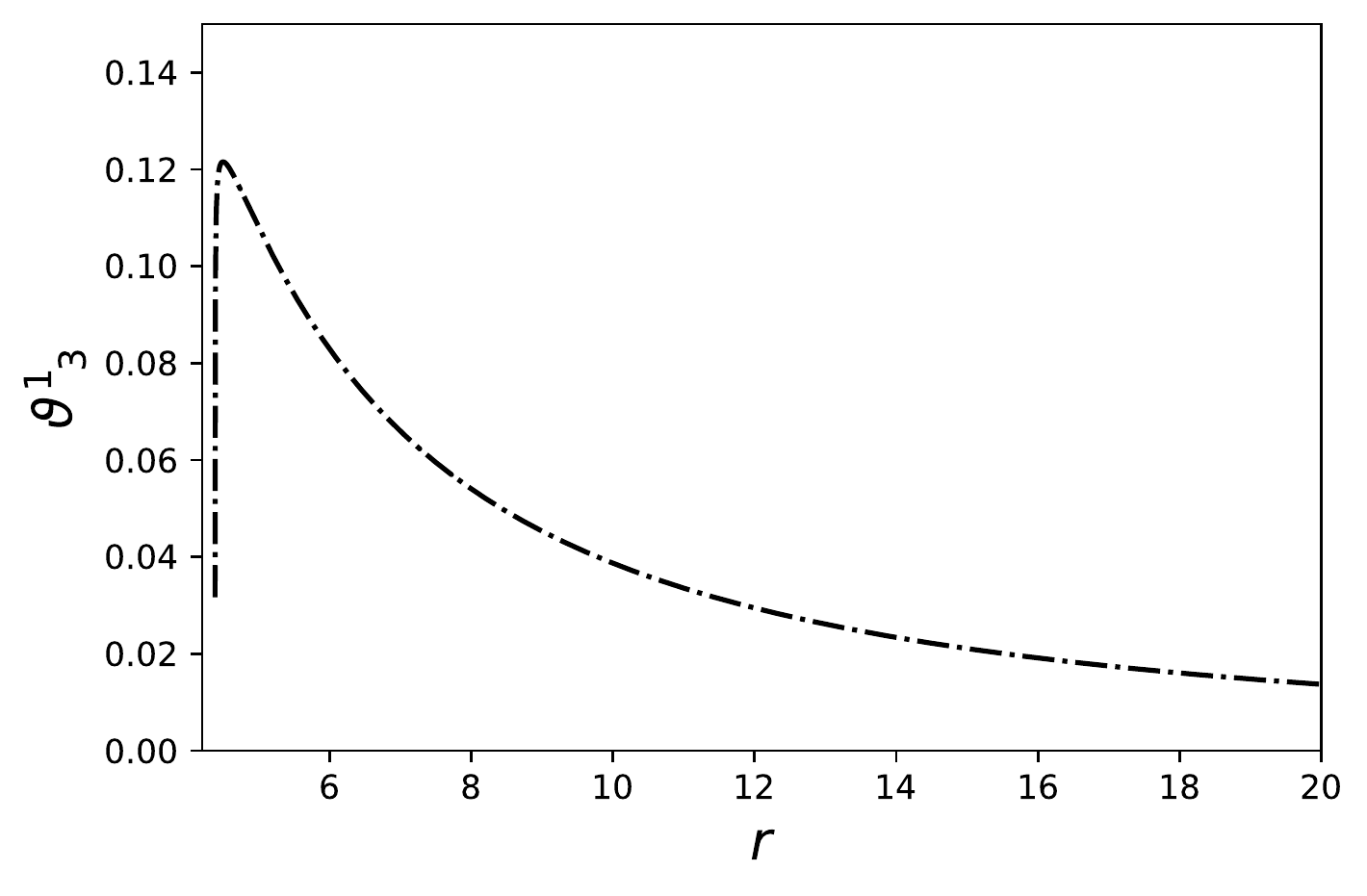}{\label{fig:f}} }}
    \caption{ The local Lorentz transformations and the local Wigner rotation for the co-rotating circular orbits as a function of $r$ with the parameters $r_s = 3$ and $a = 0.1$}
\end{figure}

For the hovering observer in circular geodesics, the non-zero local Lorentz transformation are determined by Eqs. (\ref{eq:llt}) and (\ref{eq:lltt}) and, in this case, given by
\begin{align}
    & \lambda^{0}_{\  1} = \frac{-r_sf(r)(u^t + \frac{1}{2}a u^{\phi}) - \frac{r^3_s a^2}{r^4}(u^t - a u^{\phi})}{2(r^2f(r) + r^2_sa^2/r^2)} \\
    & \lambda^{1}_{\  3} = \frac{1}{\sqrt{r^2f(r) + r^2_sa^2/r^2}}\Big( \frac{u^t r_s a}{2 r^2} - u^{\phi}\Big(\frac{r_s^2 a^2}{2 r^3} - r f(r)\Big)\Big),
\end{align}
which for $a = 0$ reduces to the local Lorentz transformation in the Schwarzschild spacetime \cite{Lanzagorta}. Thus, the local Wigner rotation is determined by Eq. (\ref{eq:lwr}). In Figs. \ref{fig:e} and \ref{fig:f}, we plotted the local Lorentz transformations and the local Wigner rotation, respectively, for the co-rotating circular orbits as a function of $r$ with the parameters $r_s = 3$ and $a = 0.1$ in some units of distance. For instance, $r_s = 3\text{ km}$ is the Schwarzschild radius of the Sun. %Therefore, each point represents the value of local Lorentz transformations and the local Wigner rotation for circular orbits with a specific radius $r$ with parameters $r_s = 3$ and $a = 0.1$.
The stable circular geodesics can be obtained by requiring that the argument of the square root of Eqs. (\ref{eq:ut}) and (\ref{eq:uphi}) must be greater than or equal to zero, which can be solved numerically for each value of $r_s$ and $a$. In this case, for the co-rotating circular orbits, the inner most stable geodesic is characterized by $r \simeq 1.4604 r_s$, which agrees with the fact that, for co-rotating circular orbits, the ISCO is closer to $r_s$ than in the Schwarzschild case \cite{Barden}. While for the counter-rotating circular orbits, the inner most stable geodesic is characterized by $r \simeq 1.53755 r_s$. Besides, from Fig. \ref{fig:f}, one can see that for $r \to \infty$, the local Wigner rotation tends to zero.

In Fig. \ref{fig:machrot}, we represent a spin-$1/2$ quanton in a `astronomical' Mach-Zehnder interferometer. The physical scenario consists of two beam splitters $BS1$ and $BS2$, a phase shifter, which gives a controllable phase $\Upsilon$ to the path $-$, and two detectors $D \pm$. The paths consist of a co-rotating and a counter-rotating circular geodesic centered in a stationary and axisymmetric body of mass $M$ and angular momentum $\mathcal{J}$. The initial state of the quanton, before the $BS1$, is given by $  \ket{\Psi_{i}} = \ket{p_i} \otimes \ket{\tau_i } = \frac{1}{\sqrt{2}}\ket{p_i} \otimes (\ket{\uparrow} + \ket{\downarrow})$, with the local quantization of the spin axis along the 1-axis. Right after the $BS1$, the state is
\begin{align}
    \ket{\Psi} & = \frac{1}{2}(\ket{p_{+};0} + i\ket{p_{-};0})  \otimes  (\ket{\uparrow} + \ket{\downarrow}) \label{eq:state2},
\end{align}
where $\phi = 0$ is the coordinate of the point where the quanton was putted in a coherent superposition in opposite directions with constant four-velocity $u^a_{\pm} = (e^{0}_{\ \mu} u^{\mu}_{\pm}, 0, 0, e^{3}_{\ \mu} u^{\mu}_{\pm})$.  Here, we assumed that the beam-splitters $BS1$ and $BS2$ do not affect the spin degree of freedom. However, it is possible to consider the beam-splitter as a Stern-Gerlach apparatus, such that the spin d.o.f will gets entangled with momentum degree of freedom. After some proper time $\tau = \frac{d \tau}{d \phi} \Phi$, the particle travelled along its circular paths and the spinor representation of the finite Wigner rotation due only to gravitational effects is given by
\begin{align}
    D(W(\pm \Phi)) = e^{- \frac{i}{2}\sigma_2 \varphi_{\pm}}, \ \ \label{eq:wigrot}
\end{align}
where $\varphi_{\pm} = \Theta_{\pm} - \Phi_{\pm}$, with $\Phi_{\pm} = \pm \Phi$ and $ \Theta_{\pm} = \vartheta_{31}(\pm)\frac{d \tau_{\pm}}{d \phi_{\pm}} \Phi_{\pm}$. Since $\vartheta^{1}_{\ 3}(x)$ is constant along the path, the time-ordering operator is not necessary. Therefore, the state of the quanton in the local frame at point $\phi = \pi$ before $BS2$ is given 
\begin{align}
U(\Lambda)\ket{\Psi}  =& \frac{1}{2} \ket{p_{+};\pi} \otimes e^{- \frac{i}{2}\sigma_y \varphi_+}(\ket{\uparrow} + \ket{\downarrow}) \nonumber \\
& + \frac{i e^{i \Upsilon}}{2}\ket{p_{-};\pi}\otimes e^{-\frac{i}{2}\sigma_y  \varphi_-}(\ket{\uparrow} + \ket{\downarrow}),  \label{eq:state1}
\end{align}
which, in general, is an entangled state. Besides, we choose to maintain the notation $\ket{p_{\pm}}$ for the momentum states before $BS2$, because the velocity of the quanton is constant along the paths. The detection probabilities corresponding to Eq. (\ref{eq:state1}), after the $BS2$, are given by
\begin{align}
    P_{\pm} = \frac{1}{2}\Big(1 \mp \cos \Big(\frac{\varphi_+ - \varphi_-}{2} \Big) \cos \Upsilon \Big).
\end{align}

\begin{figure}[t]
\centering
\includegraphics[scale=0.6]{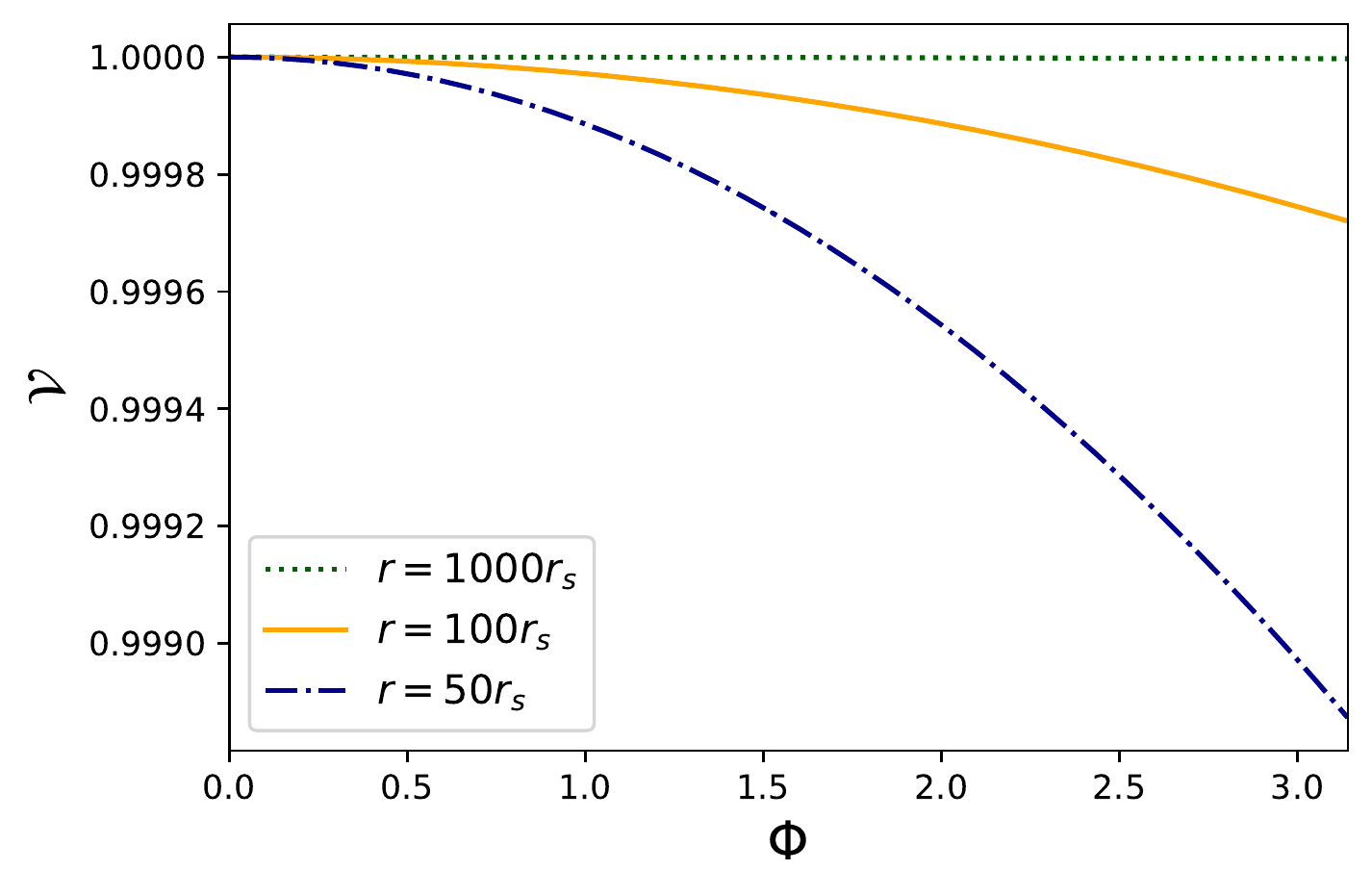}
\caption{The evolution of the visibility (or internal quantum coherence) along the `astronomical' Mach-Zehnder interferometer, since $\tau \propto \Phi$, for different values of $r$ with $r_s = 3$ and $a = 0.1$}
\label{fig:visib}
\end{figure}

When the controllable phase shift $\Upsilon$ is varied, the probabilities $P_{\pm}$ oscillate with amplitude $\mathcal{V}$ which is called interferometric visibility, and can be calculated using \cite{Zych}:
\begin{align}
    \mathcal{V} & = \abs{\bra{\tau_i} e^{-\frac{i}{2}\sigma_y (\varphi_+ - \varphi_-)} \ket{\tau_i}} \label{eq:visibi} \\
    & = \abs{ \cos \Big(\frac{\varphi_+ - \varphi_-}{2} \Big)},
\end{align}
where $\ket{\tau_i} = \frac{1}{\sqrt{2}}(\ket{\uparrow} + \ket{\downarrow})$ is the initial state of the clock. Besides, from the Eq. (\ref{eq:visibi}), one can see that the visibility will depend only in the initial state before $BS1$ and the evolution of this state between $BS1$ and $BS2$. Without the entanglement between the internal degree of freedom and the momentum, the expected visibility is always maximal, i.e., $\mathcal{V} = 1$. The introduction of the internal degree of freedom and its entanglement with the momentum, due to the fact that the Wigner rotation depends on the momentum of the particle, results in a change in the interferometric visibility. In this case, the difference of the Wigner rotation for each path can be attributed to the difference $\vartheta^{1}_{\ 3}(+) \tau_+ - \vartheta^{1}_{\ 3}(-)\tau_-$, which in part is due to the gravitomagnetic clock effect. In Fig. \ref{fig:visib}, we plotted the `evolution' of the visibility along the interferometer, since $\tau \propto \Phi$, for different values of $r$ with $r_s = 3$ and $a = 0.1$. We can see that for large $r$, the effect on the interferometric visibility is tiny, since the spacetime effect does not strongly couples the momentum with the spin, and the states of the spin for the two paths are almost indistinguishable. However, as the circular orbit gets closer to $r_s$, the effect on the visibility gets more noticeable, since the visibility can reach zero, which means that the which-way information is accessible in the final state of the clock (spin). Actually, as pointed out in Ref. \cite{Costa}, a clock with a finite dimensional Hilbert space has a periodic time evolution, and thus it is expected that the visibility oscillates  periodically as a function of the difference of the proper times elapsed in the two paths. This will also happen here if we let the quanton revolving along the superposed orbit long enough. Besides, it is expected that the same effect occurs in the Kerr's spacetime. Finally, it's worth mentioning that, if the beam splitters entangles the momentum and spin, then the visibility after $BS1$, in principle, is zero because the which-path information is encoded in the spin d.o.f. Therefore, the action of $U(\Lambda)$ in this case will be to degrade the entanglement between both degrees of freedom, and increase the quantum coherence of both degrees of freedom between the (BS1) and (BS2), since, in this scenario, entanglement and coherence are complementary quantities, as already shown by us in \cite{Jonas}.

%---------------------
\section{Conclusions}
\label{sec:con}
In this article, we presented a scenario in which a spin-$1/2$ quanton goes through a superposition of co-rotating and counter-rotating geodetic circular paths, which play the role of the paths of a Mach-Zehnder interferometer in stationary and axisymmetric spacetimes. Since the spin of the particle plays the role of a quantum clock, as the quanton moves in a superposed path it gets entangled with the momentum (or the path), and this will cause the interferometric visibility (or the internal quantum coherence) to drop, since, in stationary axisymmetric spacetimes, there is a difference in proper time elapsed along the two trajectories.  However, as showed here, the proper time of each path will couple to the corresponding local Wigner rotation, and the effect in spin of the superposed particle will be a combination of both. Besides, we discuss a general framework to study the local Wigner rotations of spin-$1/2$ particles in general stationary axisymmetric spacetimes for circular orbits. In addition, we did not took into account the spin-curvature coupling, which need to be considered when investigating spinning particles in the case of supermassive compact objects and/or ultra-relativistic test particles.

Finally, it is worth mentioning another interesting physical scenario. Instead of considering a superposed particle in a co-rotating and a counter-rotating circular orbit, as we did here, it is possible to consider a source that emits two spin-$1/2$ particles, maximally entangled on the spin states, that go in opposite circular geodetic motion. Since the momentum of each particle gets entangled with the respective internal degree of freedom through the local Wigner rotation of each path, then, since entanglement is monogamous \cite{Winter}, the initial spin-spin entanglement has to be redistributed around as spin-momentum entanglement of each particle.
  
\begin{acknowledgments}
This work was supported by the Coordena\c{c}\~ao de Aperfei\c{c}oamento de Pessoal de N\'ivel Superior (CAPES), process 88882.427924/2019-01, and by the Instituto Nacional de Ci\^encia e Tecnologia de Informa\c{c}\~ao Qu\^antica (INCT-IQ), process 465469/2014-0.
\end{acknowledgments}

%--------------------

\end{document}